\documentclass[aps,twocolumn,showpacs,prb]{revtex4} 
\usepackage{epsfig}
\usepackage{graphicx}
\def\vareps{\varepsilon}
\newcommand{\ignore}[1]{}
\begin{document}

\title{ Macroscopic vs. microscopic models of circular Quantum Dots} 
\author{J. Martorell}
\affiliation{Dept.
  d'Estructura i Constituents de la Materia, Facultat F\'isica,\\
   University of Barcelona, Barcelona 08028, Spain
}
\author{D. W. L. Sprung}
\affiliation{
  Department of Physics and Astronomy, McMaster University\\
  Hamilton, Ontario L8S 4M1 Canada
}
\date{\today}
\begin{abstract}
The confinement mechanism of electrons in gated circular quantum dots 
is studied in a sequence of models, from self-consistent 3D Hartree 
calculations to the semiclassical model of Shikin et al. Separation 
of the vertical from transverse confinement allows accurate 2D 
Hartree calculations. For moderate size dots containing $\sim 50$ electrons, 
2D Thomas-Fermi can also be accurate. Finally, a Shikin type 
parameterization of the electron density allows quantitative study of 
the importance of the many contributions to the confinement potential. 
\end{abstract}
\maketitle
\section{Introduction}

Quantum dots defined by electrostatic confinement continue to be an 
important tool for exploring the physics of electrons on the 
mesoscopic scale. Numerous Coulomb blockade experiments have provided 
information on level spectra and capacitances. At present, more 
detailed knowledge is becoming available, since new experimental 
methods allow mapping of the charge distribution inside such dots and 
even the details of the individual electron wavefunctions can be 
resolved \cite{CSS00,CST02,CSG03,PKI04,KPI05}. 

The most accurate electron wave functions are provided  by 3D  
Poisson-Schr\"odinger simulations \cite{Kumar90,JL94,Sto96}. 
These methods are highly developed but are computer intensive and 
therefore not well suited for systematic studies over a wide range of 
parameter values \cite{GMI99}. Furthermore, numerical results are 
seldom conducive to developing intuition about the importance of 
specific parameters. In parallel with these 3D approaches, a number 
of simpler two dimensional (2D) models have been developed 
\cite{MHI93,FV94,WW95,JKW97,HW99,RM02}. One of their key assumptions 
is strict confinement of the electron gas to a plane (2DEG), and 
often the transverse confining potential in that plane is taken 
arbitrarily to be a parabola or other simple analytic shape. Energy 
density functionals have been used to further simplify the 
computations while still including the main terms in the Hamiltonian. 
Free parameters available in such models can be adjusted to fit the 
data from the experiment being analyzed. The advantage of 2D models 
is that they require much less numerical effort, simplify the 
interpretation of results and allow systematic analysis. On the other 
hand, empirical parameters of these models may conceal or distort 
some aspects of the physics at play. 

Our primary objective is to connect these two approaches in a detailed 
and quantitative manner. To do so with good accuracy and a reasonable 
amount of computation, we restrict our attention to axially symmetric 
dots, which has the advantage of reducing the number of degrees of 
freedom.

The article is organized as follows: in Section II we describe the 
specific heterostructure layout that will be studied. We construct 
the total potential acting on the electrons, including the Coulomb 
interaction (direct and mirror terms),  and the electrostatic 
confining potential due to the gates and donors. The resulting 
Hamiltonian is solved numerically in the 3D Hartree approximation. By 
varying the gate voltage we present results for dots of different 
sizes. Our first step towards simpler models with a more transparent 
interpretation is the factorization {\it ansatz} between longitudinal 
and transverse components of the electron wavefunctions. This 
describes the dot in an effective 2D Hartree approximation. We end 
this section by showing that the predictions of this {\it ansatz} are 
in very good agreement with the full 3D Hartree calculations. 

Section III is devoted to the construction of effective confining 
potentials and coulomb interactions for use in 2D models, which 
incorporate finite thickness effects. We show in particular that  
there is  a substantial difference at short distances between the 
effective Coulomb interaction and the $1/r$ law. In Section IV we 
define an energy density functional based on the Thomas-Fermi (T-F) 
approximation for the kinetic energy and, via a variational 
prescription, find equations to determine the corresponding electron 
density, along with the coulomb, confinement and kinetic energies. 

In Section V, we show how to improve on the semi-classical model of 
Shikin {\it et al.} \cite{Shik89,Shik91}. We do so by taking into 
account the semiclassical kinetic energy, the mirror coulomb terms 
and the role of finite thickness of the 2DEG on the Coulomb 
interaction. We replace the Thomas-Fermi densities in the variational 
equations by the simple parametrization first introduced by Shikin 
{\it et al.} \cite{Shik89,Shik91} and later extended by Ye and 
Zaremba \cite{YZ94}. We show that after these simplifications, the 
numerical effort is reduced to a minimum: solving a very simple 
nonlinear equation for the dot radius $R$. The resulting 
predictions for bulk properties such as the number of electrons, root 
mean square radius, kinetic, confining and Coulomb energies are 
quantitatively accurate. 
Finally, by restricting the parametrized density to the simplest 
Shikin form, but keeping all relevant potential contributions, we 
show that the dot size systematics can be obtained from an analytic 
expression where each term has a clear interpretation and expresses 
the physics quantitatively.

\section{Microscopic models: Poisson- Schr\"odinger.}

\subsection{The heterostructure}                  
\begin{figure}[htpb]                 
\resizebox{8cm}{!}{\includegraphics{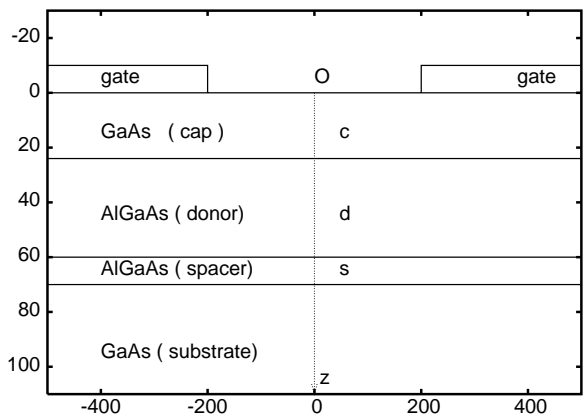}}
\caption{ Section of heterostructure layout. Lengths in nm.}
\label{figl01} 
\end{figure}

Fig. \ref{figl01} shows a section of the heterojunction that we will 
use for reference. It consists of planar layers grown on a GaAs 
substrate which we model as infinitely thick and free of acceptors. 
From bottom to top there are 1) an AlGaAs spacer layer of thickness 
$s$ , 2) an AlGaAs donor layer of thickness $d$, uniformly doped with 
donor concentration $\rho_d$, 3) a GaAs cap layer, undoped and of 
thickness $c$. The cap layer is covered by a metallic gate except for 
a circular opening of radius $S_0$ exposing the GaAs cap layer. 
Since our main purpose is to compare the predictions of 
several models, rather than to produce an accurate description of a 
specific experiment, we introduce several simplifying assumptions 
that will make our calculations less cumbersome, but which should 
not affect the physics to be explored. The first of these 
is the gate geometry just described. Using expressions to be 
given below, we have checked that for rings of a few hundred 
nanometers width, and same inner radius $S_0$, the potential due to 
the gates is practically identical to that of the simplified gate 
used here, at the location of the dot. Also for simplicity we neglect 
the differences in dielectric constant $\vareps_r$ between the 
different materials, and use a common energy-independent effective 
mass for spacer and substrate. 

Since in an earlier work \cite{MWS94} we modelled quantum wires using 
similar microsopic models we have adopted the same geometrical and 
physical parameters for the heterostructure: $c= 24 $ nm, $d= 36 $ 
nm, $s = 10 $ nm, and $\rho_d = 6. 10^{-4}$ nm$^{-3}$, $\varepsilon_r 
= 13.1$, $m^*/m_e = 0.067$. These values are taken from a model 
heterostructure considered by Laux {\it et al.} \cite{Laux88} in 
their pioneering work on wires. We have discussed how to deal 
with 
partially ionized donor layers in previous work \cite{MWS94}. Here 
for simplicity we will assume the donors are totally ionized. (This 
is often achieved in experiments by illuminating the sample.) 
Finally, we assume that the Fermi level is pinned at the exposed 
surface. This again would depend on how the sample has been 
processed, but for modelling purposes it has the advantage of being a 
definite prescription, which has been used frequently in a 
variety of problems \cite{JL94,Davies88,LDK93,JUYB04,YBB01}. It is 
also a useful reference  for other models of surface states 
\cite{DLS95,RD03,PIK02,FIM02}. As indicated in Fig. \ref{figl01}, we 
choose the origin of coordinates at the center of the exposed 
surface, and the $z$ axis orthogonal to the layers. Vectors from that 
axis, in planes parallel to the layers, are denoted by letters ${\vec 
r}$ or ${\vec s}$.  Following  \cite{MWS94}, we construct the 
electrostatic potential acting on the electrons by adding the 
contributions from a) the gate, b) the ionized donor layer, and c) 
electrons in the dot. The potential due to a metallic gate, at 
voltage $V_0$ and with a circular hole of radius $S_0$, can be 
derived from expressions given in Sect. 3.6 of \cite{Davies88}: 
 \begin{equation}                       
eV_g({\vec r},z) = {1 \over {2\pi}} \int \ d{\vec s}\ eV_g({\vec
s},0)\ {{\vert
z \vert} \over {(z^2 +\vert {\vec r} - {\vec s} \vert^2 )^{3/2}}} \ ,
\label{eq:le1}
\end{equation}
where $e$ is the electron charge ( $e<0$.)
Using the boundary condition for Fermi level pinning, 
 \begin{equation}                       
eV_g({\vec s},0) = eV_0 \ \Theta( s - S_0) \ ,
\label{eq:le2}
\end{equation}
one finds, for $z>0$, 
 \begin{eqnarray}                       
eV_g(r,z) &=& eV_0 {2 z \over \pi}  \int_{S_0}^{\infty} s \ ds \
E\left( {{4 r s} \over {z^2 +(r+s)^2}}\right)  \nonumber \\
&& \qquad
{1 \over {z^2 +(r-s)^2}} {1 \over {(z^2 +(r+s)^2)^{1/2} }}
 \ . \label{eq:le3}
\end{eqnarray}
Here $E(k^2)$ is the complete elliptic integral of the second kind as 
defined in \cite{AS64}. To this one must add the electrostatic 
potential due to the donor layer, including the corresponding mirror 
term which maintains the boundary condition at the surface. As in 
\cite{MWS94}, in the substrate and spacer layers this contribution 
reduces to a constant additive term: 
 \begin{equation}                       
eV_d (z ) = -{e^2 \over \varepsilon} \rho_d\ d ( c +{d \over
2}) \ \ .
\label{eq:le4}
\end{equation}
The third contribution is due to the Coulomb interaction between the 
electrons. Including the mirror terms it is: 
 \begin{eqnarray}                       
e V_e({\vec r},z; {\vec r}^\prime,z') &=& {e^2\over \varepsilon_r}
\bigg[ {1 \over {{\sqrt{\vert {\vec r}- {\vec r}^\prime\vert^2+(z-z')^2}  }}}
\nonumber \\
&& \qquad -{ 1 \over {{\sqrt{\vert {\vec r} - {\vec r}^\prime\vert^2+(z+z')^2} }}}
\bigg]
\label{eq:le5}
\end{eqnarray}
Appropriate band offsets in the AlGaAs layers,
$eV_{bo}= 0.23 $ eV, are included, but for brevity we do not write them in  
the expressions below. Choosing the Fermi level as the origin 
of energies and  denoting by $e V_s$ the binding energy of the 
surface states with respect to the conduction band edge, the latter 
will be located at: 
 \begin{equation}                       
eV_{g+d}(r,z)=eV_s + eV_g(r,z) + eV_d(z) \ .
\label{eq:le6}
\end{equation}
We have taken $e V_s = 0.7 $ eV.\\ 
The Hamiltonian for $N$ electrons in the dot is:
 \begin{eqnarray}                       
H &=& \sum_{i=1}^N \left( {{\hat p}_i^2 \over {2m^*}} + eV_{g+d}(r_i,z_i)
\right) \nonumber \\ && \qquad
+ \sum_{i<j=1}^N eV_e({\vec r}_i,z_i;{\vec r}_j,z_j) \ ,
\label{eq:le7}
\end{eqnarray}
which we solve in the Hartree approximation. 

\subsection{3D Hartree approximation}             
We take a trial product wave function  
 \begin{equation}                       
\Psi({\vec r}_1,z_1; ....; {\vec r}_N,z_N) = \prod_{i=1}^N
\psi_i ({\vec r}_i,z_i)~, 
\label{eq:le8}
\end{equation}
and minimize the trial energy
\begin{equation}                        
E_H =  <\Psi| H | \Psi>  \ ,
\label{eq:le9}
\end{equation}
under constraints that impose 
orthonormality of the orbitals $\psi_i$. To guarantee equilibrium 
with the surface states, all single particle states of negative energy 
are occupied. The variational method gives 
\begin{eqnarray}                      
\bigg[ {{\hat p}_i^2 \over {2m^*}} &+& e V_{g+d} (r_i,z_i) + 
U_H({\vec r}_i,z_i) 
 \bigg] \psi_i = \varepsilon_i \psi_i({\vec r}_i,z_i) \nonumber \\
 U_H({\vec r},z) &=& \int d{\vec r}^\prime dz'\ eV_e({\vec r},z;
{\vec r}^\prime,z')\ \rho({\vec r}^\prime, z')
\label{eq:le10}
\end{eqnarray}
with
\begin{equation}                        
\rho({\vec r},z) = \sum_{i=1}^N |\psi_i({\vec r},z)|^2 \ .
\label{eq:le11}
\end{equation}
We will consider only dots with an even number of electrons. In that 
case the ground state solution has axial symmetry, so that 
 \begin{equation}                       
\rho({\vec r},z) = \rho(r,z) \quad , \quad U_H({\vec r},z ) = U_H(r,z) \ ,
\label{eq:le12}
\end{equation}
and the wavefunctions factorize as 
 \begin{equation}                       
\psi_{nl}({\vec r},z) = {{u_{nl}(r,z)}\over {\sqrt{r}}}\ {e^{i l 
\theta}\over {\sqrt{2\pi}}}~. 
 \label{eq:le13}
\end{equation}
We have omitted for simplicity the spin component of the 
wavefunction. Furthermore, eq. \ref{eq:le10} simplifies to a two-dimensional 
Schr\"odinger equation  
 \begin{eqnarray}                       
\varepsilon_{nl} u_{nl}(r,z)  &=&
-{\hbar^2\over{2m^*}}\left({{\partial^2}\over 
{\partial r^2}} +{{\partial^2}\over{\partial z^2}}
\right) u_{nl}(r,z) +  \nonumber \\ 
\bigg( eV_{g+d}(r,z) &+& U_H(r,z)+{\hbar^2\over{2m^*}}{{l^2 - 
{1\over 4}}\over r^2} \bigg) u_{nl} \ .
 \label{eq:le14}
\end{eqnarray} 
As usual, we have solved the system formed by eqs. \ref{eq:le10} and 
\ref{eq:le14} by numerical iteration. Before presenting results, we 
explore factorized approximate solutions of eq. \ref{eq:le14} .

\subsection{Factorization ansatz}                 
It is well known that large confining energies in the longitudinal 
direction support the validity of approximations where the $z$ 
dependence in the wavefunctions is factored out from the other 
degrees of freedom. Guided by the success of the {\it ansatz} 
employed for linear wires \cite{MWS94},  we use it here for dots. We 
begin by approximating the two potential terms in  eq. \ref{eq:le14} 
as 
 \begin{eqnarray}                         
e V_{g+d}(r,z) &+& U_H({\vec r},z) = \nonumber \\
 &\simeq& e V_{g+d}(0,z) + {\overline {\Delta U}}_c(r) +
\overline{U}_H({\vec r})
\label{eq:le15}
\end{eqnarray}
where the effective two-dimensional confining, ${\overline{\Delta 
U}_c}$, and Hartree, $\bar{U}_H$, potentials are defined as: 
 \begin{eqnarray}                         
\overline {\Delta U}_c(r) &=& \int dz A^2(z) ( eV_{g+d}(r,z) -
eV_{g+d}(0,z) )
\nonumber \\ 
\overline{U}_H({\vec r}) &=&  \int dz \ A^2(z) U_H({\vec r},z)
\label{eq:le16}
\end{eqnarray}
with a weight function $A^2(z)$ to be specified below. 
Formally the range of integration over $z$ extends from $-\infty$ to 
$+\infty$, but in practice the $A(z)$ are non-zero only in the spacer 
and the substrate. Making these approximations in eq. \ref{eq:le10} 
gives 
 \begin{equation}                          
\left[ {{\hat p}^2\over {2m^*}} + eV_{g+d}(0,z) +\overline{\Delta
U}_c(r)
+ \overline{U}_H({\vec r}) \right] \psi_i^{(a)} = \varepsilon^{(a)}_i
\psi_i^{(a)} 
\label{eq:le17}
\end{equation}
that has solutions 
 \begin{equation}                       
\psi_i^{(a)}({\vec r},z) = A(z) \phi_i({\vec r})~,
\label{eq:le18}
\end{equation}
with separately normalized $A(z)$ and $\phi_i$, which satisfy
Schr\"odinger equations with eigenvalues $E_z$ and
$e_i$ 
 \begin{eqnarray}                       
\left[ {{\hat p}_z^2\over {2m^*}} +  eV_{g+d}(0,z) \right] A(z) &=&
E_z A(z) \label{eq:le19a}  \\
\left[ {{\hat p}_{\perp}^2 \over {2m^*}} + \overline{\Delta U}_c(r) +
\overline{U}_H({\vec r}) \right] \phi_i({\vec r}) &=& e_i
\phi_i({\vec r}) 
\label{eq:le19b}
\end{eqnarray}
Note that we have chosen the previously unspecified weight function 
$A(z)$ to be the common longitudinal component of the wavefunctions 
in \ref{eq:le18}. We then have 
 \begin{eqnarray}                       
\varepsilon_i^{(a)} &=& E_z + e_i  \nonumber \\
E_H^{(a)} &=&  N E_z + \sum_{i=1}^N < \phi_i\ \vert {{\hat 
p}_{\perp}^2 \over {2m^*}} +
\overline{\Delta U}_c + {1 \over 2} \overline{U}_H\ \vert \phi_i >
\label{eq:le20}
\end{eqnarray}
The numerical process starts by constructing $eV_{g+d}(0,z)$ and
solving eq. \ref{eq:le19a}. After that $A(z)$ is fixed and we solve the 
two-dimensional Hartree problem iteratively, constructing the 
potentials from  eqs. \ref{eq:le16}, and solving equation 
\ref{eq:le19b} for the $\phi_i({\vec r})$, $i=1, ... N$. 
For dots with an even number of electrons, axial 
symmetry  allows separation of radial and angular 
factors in the $\phi_i({\vec r})$, as done in eq. \ref{eq:le13}, and 
eq. \ref{eq:le19b} reduces to a one-dimensional 
equation for the radial components, which is straightforward to 
solve.

\begin{figure}[htpb]                 
\resizebox{8cm}{!}{\includegraphics{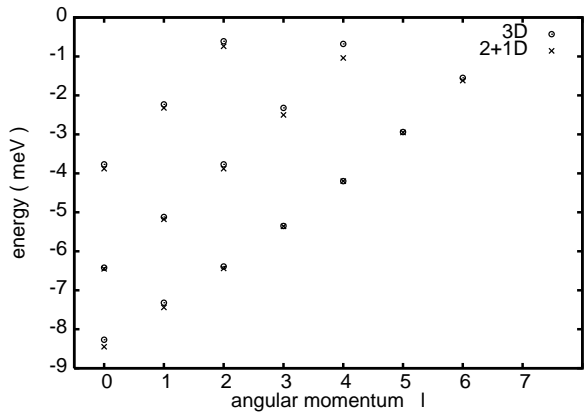}}  
\caption{ Energies of single electron levels, corresponding 
to $V_0 = - 1.27 $ eV and $S_0 = 210$ nm. Circles: 3D Hartree, $x$ symbols: 
factorization ansatz.} 
\label{figl02}
\end{figure}

\subsection{Results}                    
In Fig. \ref{figl02} we show, for $ N= 54$ electrons, the orbital 
energies as a function of angular momentum. One sees that 
results using the factorization ansatz (labelled 2+1 D) are in 
excellent agreement with the exact Hartree (labelled 3 D). The 
various bulk energies are also accurately predicted by the 
factorization ansatz: To give two examples: {\it 1)} when $V_0= -1.32 
$  V, we find a ground state with $N=40$ electrons. The kinetic, 
confining, Coulomb and total energies are found to be  $0.720\, (0.729) $  
eV, $-3.263\, (-3.272) $ eV, $1.215\, (1.214) $ eV and  $-1.327\, (-1.329) $ 
 eV respectively. 
The quantities in parentheses correspond to the ansatz. 
{\it 2)} when $V_0 = -1.27  $ V, 
$N= 54$,  the respective values are $0.972\, (0.985)$, $ -5.076\, (-
5.095)$ , $ 1.947\, (1.949)$ and $-2.160\, (-2.161) $  eV. 

\begin{figure}[htpb]                 
\resizebox{8cm}{!}{\includegraphics{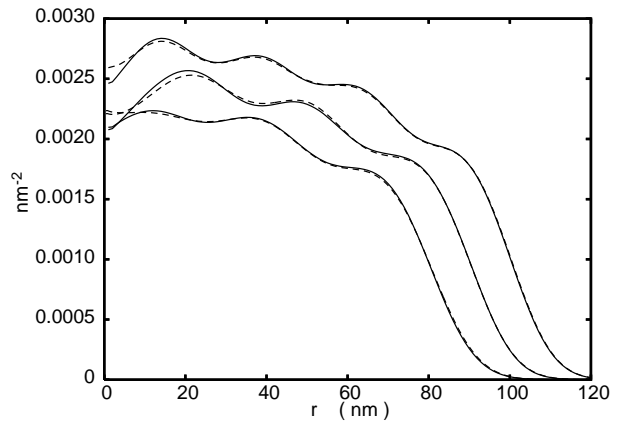}}
\caption{ 2D electron densities $\sigma(r)$ computed in the Hartree 
approximation (continuous lines) and the factorization ansatz (dashed 
lines). $N= 40,\, 54$ and $72$ electrons.} 
 \label{figl03} 
\end{figure}

A 3D plot of the electron density $\rho(r,z)$ of the Hartree 
approximation shows that the dot extends about 20 nm along the $z$ 
axis with a smooth variation very similar to the square of the Airy 
function. This will be exploited in Appendix A. Along $r$ the 
density shows oscillations due to filling of the various orbitals. 
This is displayed in Fig. \ref{figl03} where we have 
plotted: 
 \begin{equation}                       
\sigma_H(r) \equiv \int \rho(r,z) \ dz. 
\label{eq:le21}
\end{equation}
for $N = 40,\, 54,\, 72$, and the corresponding quantity in the 
factorized approximation, 
 \begin{equation}                       
\sigma^{(a)}_H(r) = \sum_{i=1}^N |\phi_i(r)|^2  \ .
\label{eq:le22}
\end{equation}

Again, the agreement between the Hartree (3D) and the factorized 
densities is very good. A plot of the same densities of Fig. 
\ref{figl03} on a logarithmic scale, shows that the agreement for the 
large $r$ is excellent down to values as low as 
$ \sigma_H(r) \simeq 10^{-6}$ nm$^{-2}$. 
This suggests that the factorization ansatz might be accurate in 
studies of tunneling across a ring of finite width.  The root mean 
square radii are also accurately predicted: for $N=40$ electrons we 
find $ 57.5\, (57.6) $  nm, whereas for $N=54$ the two values agree to 
three digits,  $63.5 $  nm.  
 
\begin{figure} [htpb]                
\leavevmode
\epsfig{file=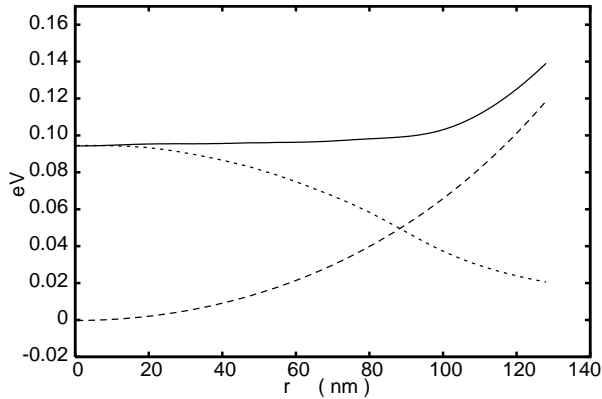,width=8cm}
\caption{ 2D potentials ${\bar {\Delta U}}_c(r)$ (dashed line), 
${\bar U}_H(r)$ (dotted line) and their sum (continuous line). 
$N = 54$, corresponding to $V_0 = - 1.27 $ eV and $S_0 = 210$ nm.} 
 \label{figl04}
\end{figure}

Fig. \ref{figl04} shows the Hartree and confining potentials. Both are 
nearly parabolic for the radii corresponding to the area occupied by the dot 
($r \le 100$ nm), and their sum is almost constant in this range of $r$. 
We also find that the kinetic energy is small on the scale of the 
figure. These findings will be used in Sections IV and V to support, and 
improve on, the macroscopic model of Shikin {\it et al.} 
\cite{Shik91}. 

\section{Potentials in the 2D limit}

Here, our aim is to develop the relation between the 3D calculations 
of the last section, and purely 2D models.
We make use of a separable expansion of the confining potential 
developed in Appendix B: 
 \begin{equation}                        
eV_g (r,z) = eV_g(0,z) \ + \ {1\over 2 } k_c(z) r^2 \ + \ {1\over 2} k_4(z)
r^4 \ ...
\label{eq:le23}
\end{equation}
Expressions for the coefficients are given in eq. \ref{eq:bp5}. 
To have a workable expression for $\overline {\Delta 
U}_c(r)$ defined in eq. \ref{eq:le16}, we approximate the integral 
over $z$ by replacing the exact $A(z)$ by Airy functions (see 
Appendix A).  This leads to  
 \begin{equation}                          
{\overline {\Delta U}}_c(r) \simeq U_{2D,c}(r)  \equiv {1\over 2} 
{\bar k}_c r^2+ {1\over 2} {\bar k}_4 r^4 
\label{eq:le24}
\end{equation}
with the functions $k_c(z)$ and $k_4(z)$ replaced by their values at 
$z = {\bar z}$ given in eq. \ref{eq:ap12}. From now on we will use 
this truncated form of the confining potential, $U_{2D,c}(r)$ with 
terms parabolic and quartic in $r$. 

Guided by eq. \ref{eq:le16}, we define the effective 2D coulomb 
interaction as 
 \begin{equation}                         
{\varepsilon_r \over e^2} \ V_{2D}(|{\vec r}-{\vec r}^\prime|) = \int \int 
{{A^2(z) A^2( z') \ dz \ dz'} \over{\sqrt{|{\vec r}-{\vec r}^\prime|^2+ (z-z')^2}}} 
\ , 
 \label{eq:le25}
\end{equation}

\begin{figure}[htpb]                
\leavevmode
\epsfxsize=8cm
\epsffile{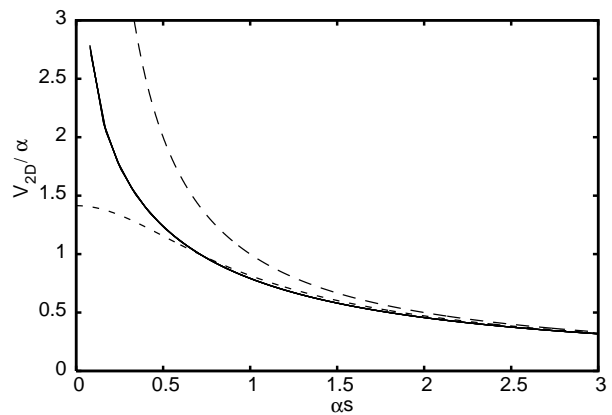}
\caption{ Effective 2D Coulomb interactions. Continuous line: effective 
interactions defined in eq. \protect{\ref{eq:le25}}. Upper long dashed 
line: $1/s$ interaction. Lower short dashed line: parametrization of 
eq. \protect{\ref{eq:le27}}  } 
 \label{figl05}
\end{figure}

We now look for a suitable approximation that simplifies the 
averaging over the $A(z)$. To do so we again use that the 
$A(z)$ are well approximated by the Airy function. With notation 
introduced in Appendix A, we write 
 \begin{equation}                         
{\varepsilon_r \over e^2} \ V_{2D}(s) = {\alpha \over 
{(Ai'(\xi_0))^4}} \int \int {{Ai^2(u) Ai^2( u') \ du \ du'} 
\over{\sqrt{s^2 \alpha^2 + (u-u')^2}}} \ , 
\label{eq:le26}
\end{equation}
with $s = |{\vec r}-{\vec r}^{ \prime}|$. In this approximation, the 
ratio $V_{2D}(s) / \alpha$ depends only on the product 
$s\, \alpha $. The parameter $\alpha$ 
contains all the information on the confining potential and on the 
specific heterostructure. We have checked that to a very good 
approximation the effective interactions obtained numerically from 
eq. \ref{eq:le25} for different values of $S_0$ and the gate 
potential $V_0$ also 
satisfy this simple relation. The continuous line in Fig. \ref{figl05} 
corresponds to the curves  $({\varepsilon_r / e^2}) \, (V_{2D}(s) /\alpha)$ 
computed from eq. \ref{eq:le25} for a range of values of 
$S_0$ from $150$ to $260 $ nm and of $V_0$ from $-1.12$ to $-1.42 $ V. 
On the scale of the figure they are all superimposed. The upper 
dashed line is for the pure $1/s$ potential and shows that the finite 
thickness effect (from averaging over $A(z)$) is substantial. For the 
simplified 2D models that we will introduce below, it will be very 
useful to have a simple analytic approximation for the effective 
interaction: empirically, we have found that the expression 
 \begin{equation}                        
{\varepsilon_r \over e^2} \ V_{2D}(s) = {1\over{\sqrt{s^2+ t^2(\alpha)}}} \ .
\label{eq:le27}
\end{equation}
with $t^2(\alpha) = 1/(2 \alpha^2)$ gives a fairly good fit to the 
long range part of the interaction, and is very simple to use. 

Its physics content is also clear: we have replaced the $(z-z')^2$ in 
the denominator of eq. \ref{eq:le25} by a constant, $t$, which plays 
the role of an effective thickness of the 2DEG. In this approximation 
the effective thickness is proportional to $\gamma^{-2/3}$, $\gamma$ 
being the slope (in $z$) of the confining potential. In the following 
we will show that this effective interaction reproduces the 
direct Coulomb potential in the plane of the 2DEG very well. 

A similar average has to 
be performed to define an effective mirror interaction. However, the 
width of the $A(z)$ functions is small compared to the value of $z+z' 
\simeq 2 z_m \equiv D$, with $z = z_m$ the maximum of $A(z)$,  and 
therefore it is a good approximation to write 
 \begin{eqnarray}                       
{\varepsilon_r \over e^2} \ V_{2D,m}(|{\vec r}-{\vec r}^{ \prime}|) &=& 
- \int \int {{A^2(z) A^2( z') \ dz \ dz'} \over 
{\sqrt{|{\vec r}-{\vec r}^{ \prime}|^2+ (z+z')^2}}} 
\nonumber \\ 
&\simeq& - {1 \over{\sqrt{|{\vec r}-{\vec r}^{ \prime}|^2+ D^2}}} 
 \label{eq:le28}
\end{eqnarray}

\section{ The 2D energy density functional.}

The Hartree approximation is expected to be valid when the number of 
electrons in the dot is not small. In that case, the kinetic energies 
involved are  small compared to confinement and coulomb \cite{SM96}, 
and a Thomas-Fermi energy density functional becomes a sufficient 
approximation when the main interest is in the bulk properties of the 
system, rather than in details of each individual electron 
wavefunction and energy.  Besides requiring much less numerical 
effort than solving eq. \ref{eq:le19b} for all the $\phi_i$, this 
simpler model will allow us to obtain several analytic approximations 
that clarify the roles of the physical parameters of the system.  

We replace the two-dimensional density constructed from the 
solutions of eq. \ref{eq:le19b} by its Thomas-Fermi approximation, 
$\sigma_{TF}(r)$. By definition the latter is to be be determined from
a variational condition on a suitably simplified expression for the
total energy. Starting from eq. \ref{eq:le20} we write 
 \begin{equation}                      
E_T = N E_z + E_{kin} + E_{con} + E_{Coul}
\label{eq:tf1}
\end{equation}
with 
 \begin{eqnarray}                      
N &=& \int  \sigma_{TF}(r) \ d{\vec r} \nonumber \\
E_{kin} &=& {\hbar^2 \over {2m^*}} \pi \int \sigma^2_{TF}(r) \ d{\vec r}
\nonumber \\
E_{con} &=& \int \overline{\Delta U}_c(r) \  \sigma_{TF}(r) \ d{\vec r}
\nonumber \\
E_{Coul} &=& {e^2 \over {2 \varepsilon_r}} \int \left(V_{2D}(|{\vec
r}-{\vec r}^{ \prime|}) + V_{2D,m}(|{\vec r}-{\vec r}^{ \prime}|)\right)
\sigma_{TF}(r)\nonumber \\ &.& \sigma_{TF}(r') \ d{\vec r} \ d{\vec
r'} \ ,
\label{eq:tf2}
\end{eqnarray}
where we have used the well known expression for the kinetic energy
$E_{kin}$ of a 2D Thomas-Fermi gas. Defining now:
 \begin{equation}                      
E(\mu) = E_T - \mu N
\label{eq:tf3}
\end{equation}
with a Lagrange multiplier $\mu$ to fix the number of electrons, the 
variational condition $\delta E(\mu)/ \delta \sigma_{TF}$ leads to 
 \begin{equation}                      
\mu = E_z + {\hbar^2 \over {m^*}} \pi \sigma_{TF}(r) +
\overline{\Delta U}_c(r) + \overline{U}_{H,TF}(r)
\label{eq:tf4}
\end{equation}
where:
 \begin{eqnarray}                      
\overline{U}_{H,TF}(r)  &\equiv& U_{2D,d}(r) + U_{2D,m} \nonumber \\ 
&=& {e^2 \over \varepsilon_r}  \int \left(V_{2D}(|{\vec r}
-{\vec r}^{ \prime}|) + V_{2D,m}(|{\vec r}-{\vec r}^{ \prime}|) \right) 
\nonumber \\
&.& \sigma_{TF}(r') \ d{\vec r'} \ ,
\label{eq:tf5}
\end{eqnarray}
These two equations define the iterative Thomas-Fermi approach. The 
process is started by solving the Schr\"odinger eq. \ref{eq:le19a} 
for $A(z)$ to determine $E_z$.  Then $\overline{\Delta U}_c$ of 
eq. \ref{eq:le24} is computed.  The choice previously made for the 
origin of energies fixes $\mu = 0$.  Then by iteratively solving eqs.  
\ref{eq:tf4} and \ref{eq:tf5} we determine $\sigma_{TF}$ (and $N$). 
Note that we do take the longitudinal 
energy $E_z$ into account in a realistic way, and therefore the 2D 
Fermi level of the dot is not $0$ but $-E_z$.  If we had taken it to 
be zero the results would be completely different.\\ We will not 
present the numerical results found in this approach, because 
they are very similar to those we will derive next, using parametrized 
analytic forms of the density.  These prove to be more convenient for 
obtaining bulk properties in analytic form. 

\section{Parametrized densities} 

\subsection{Extended Shikin model}                
The 2D Thomas-Fermi model  is based on a semiclassical approximation 
for the kinetic energy. Some time ago, Shikin {\it et al.} 
\cite{Shik91,Shik89} noticed that the dominant contributions should 
be those of the coulomb and confining potentials and proposed a more 
drastic approximation: a model where both the kinetic energy contribution 
and mirror term were neglected in writing the total energy of the 
dot. Further they assumed that a parabolic  confining potential 
 \begin{equation}                     
\overline{\Delta U}_c^{S}(r) = {1 \over 2} k r^2 \ .
\label{eq:sh1}
\end{equation}
With these simplifications they obtained an  analytic solution for the
charge distribution:
 \begin{equation}                     
\sigma_p(r) = \sigma_2\, \sqrt{ 1 - {r^2 \over R^2}} \,\, \Theta(R-r)
\label{eq:sh2}
\end{equation}
with $\sigma_2$ and $R$ given in terms of the parameters 
defining the potentials.  Subsequently, Ye and Zaremba \cite{YZ94} extended 
this  model by adding  quartic terms in the confining 
potential, and found the corresponding analytic solution for the 
electron distribution 
 \begin{eqnarray}                     
\sigma_q(r) &=& \bigg[ \sigma_2 \left( 1 - {r^2 \over R^2}\right)^{1/2} 
+ \sigma_4 \left( 1 - {r^2 \over R^2}\right)^{3/2}\bigg] \Theta(R-r)~.  
\nonumber \\
\label{eq:sh3}
\end{eqnarray}
We extend the above models by including accurate approximations for 
the contributions of the mirror potential, the thickness correction 
to the coulomb direct potential  and the kinetic energy.  We will 
show that one can still obtain analytic expressions and that our 
extension leads to results for bulk properties of the dots that 
compare very well with the Hartree calculations of the previous 
sections.\\ Let us begin by remembering that in the limit of a 
strictly two-dimensional quantum dot and a $1/r$ interaction, the 
direct coulomb potential corresponding to the parametrized densities 
above is analytic 
 \begin{eqnarray}                     
U_{2D,d}(r) &=& {{\pi^2 e^2}\over{2\varepsilon_r}} R \bigg[\sigma_2 
\left(1- {r^2\over {2R^2}}\right) \left(1- \tau_2 \left( 1 -{r^2\over 
R^2} \right )\right) 
 \nonumber \\ 
&+& {3\over 4} \sigma_4 \left(1- {r^2\over R^2}+ {3\over 8} 
{r^4\over R^4} \right)
\left(1 -\tau_4 \left( 1 -{r^2\over R^2}\right)  \right) \bigg] 
\nonumber \\ \ ,
&& {\rm when} \quad r \le R , 
 \label{eq:sh4} 
\end{eqnarray} 
and with the constants $\tau_2$ and $\tau_4$ equal to zero. 
In Appendix C we derive approximations for the direct potential 
which take into account the finite thickness of the dot using 
eq. \ref{eq:le27}. These approximations lead to the same expression, 
eq.  \ref{eq:sh4}, but with non-zero values for the constants $\tau_2$ 
and $\tau_4$, which are given in eqs. \ref{eq:cp4} and \ref{eq:cp8}.\\ 
The mirror potential can also be expressed in a simplified form, 
which is still sufficiently accurate, using eq. \ref{eq:le28} 
 \begin{eqnarray}                     
U_{2D,m}(r) &=& - {e^2 \over \varepsilon_r} \int \ d^2 r' 
{{\sigma_q(r')} \over {\sqrt{D^2 + \vert {\vec r} - 
{\vec r}^{ \prime} \vert ^2}}}   \nonumber \\ 
&\simeq& - {e^2 \over \varepsilon_r} \int \ d^2 r' {{\sigma_q(r')} 
\over {\sqrt{D^2 + {R_u^2\over 2}+ r^2}}} \nonumber \\ 
&=& -{e^2 \over \varepsilon_r} { N \over {\sqrt{L^2+ 
r^2}}} ~, 
 \label{eq:sh5} 
\end{eqnarray} 
where $L^2 = D^2 + R_u^2/2$. Note that we have replaced the $\vert 
{\vec r} -{\vec r}^{ \prime}\vert ^2$ by its average over $r'$ on a uniformly 
charged disk of radius $R_u$ that contains the dot charge. 
The corresponding power series expansion is 
 \begin{equation}                      
U_{2D,m}(r) = -{{e^2 N}\over{\varepsilon_r L}} \left( 1-
{r^2\over{2L^2}} + {3\over 8} {r^4\over L^4} + ...\right)~, 
 \label{eq:sh7}
\end{equation}
where according to eq. \ref{eq:sh3} the number of electrons is 
\begin{equation}                      
N = {2\over 3} \pi R^2  \left[ \sigma_2 + {3\over 5} \sigma_4 \right]~.
\label{eq:sh8}
\end{equation}
Finally, for the small contribution of the kinetic energy term, we 
use the truncated expansion 
 \begin{eqnarray}                    
{{\hbar^2 \pi}\over m^*} \sigma_q &=& {{\hbar^2 \pi}\over m^*} \bigg[ 
\sigma_2 \left(1-{r^2\over{ 2R^2}} - {r^4\over{8 R^4}} \right) 
\nonumber \\ 
 &+&  \sigma_4 \left( 1-{{3r^2}\over {2R^2}} + {{3 r^4}\over{8R^4}} 
\right) \bigg] \ . 
 \label{eq:sh9}
\end{eqnarray}
With the approximations described above, eq. \ref{eq:tf4} can now be 
written as 
 \begin{equation}                    
E_z +  U_{2D,c} + U_{2D,d} + U_{2D,m} 
+ {{\hbar^2 \pi}\over m^*} \sigma_q(r) = 0
\label{eq:sh10} 
\end{equation}
and should be satisfied for all $r\le R$. 
Explicit forms for each of these terms have been written in 
eqs. \ref{eq:le24}, \ref{eq:sh4}, \ref{eq:sh7} and \ref{eq:sh9} . 
Equating the coefficients at each order of the power series, we 
arrive at 
 \begin{eqnarray}                    
 -E_z &=& \sigma_2 \left[ {{\hbar^2\pi}\over m^*} + {{\pi^2 e^2 
}\over {2\varepsilon_r}} R (1-\tau_2) - {{2e^2 \pi}\over{3\varepsilon_r 
L}} R^2\right] \nonumber \\ 
 &+& \sigma_4\left[{{\hbar^2\pi}\over m^*} + {{3\pi^2 e^2 }\over 
{8\varepsilon_r}} R (1-\tau_4)  - {{2e^2 \pi}\over{5\varepsilon_r L}} 
R^2\right] \nonumber \\ 
-{1\over 2} {\bar k}_c &=& \sigma_2 \left[-{{\hbar^2\pi}\over 
{2m^*R^2}} + {{\pi^2 e^2 }\over {4\varepsilon_r R}} (3 \tau_2-1)  + {{e^2 
\pi}\over{3\varepsilon_r L^3}} R^2\right]  \nonumber \\ 
 &+& \sigma_4 \left[- {{3\hbar^2\pi}\over {2 m^* R^2}} + {{3\pi^2 e^2 
}\over {8\varepsilon_r R}} (2\tau_4-1) + {{e^2 \pi}\over{5\varepsilon_r 
L^3}} R^2\right] \nonumber \\ 
-{1\over 2} {\bar k}_4 &=& \sigma_2 \left[- {{\hbar^2\pi}\over {8m^* 
R^4}} - {{\pi^2 e^2}\over{4 \varepsilon_r R^3}}\tau_2 - {{e^2 
\pi}\over{4\varepsilon_r L^5}} R^2\right] \nonumber \\ 
 &+& \sigma_4 \left[{{3\hbar^2\pi}\over {8m^* R^4}} + {{3 \pi^2 e^2 } 
\over {64\varepsilon_r R^3}}  (3-11 \tau_4)  - {{3e^2 
\pi}\over{20\varepsilon_r L^5}} R^2\right]~,   \nonumber \\ 
\label{eq:sh11}
\end{eqnarray}
which is a system of equations in three unknowns: $\sigma_2$ , 
$\sigma_4$ and $R$. Since the equations are linear, 
one can eliminate $\sigma_2$ and $\sigma_4$ leaving 
a single nonlinear equation in $R$, which is 
then solved numerically, by Newton-Raphson. We have 
already seen that for each value of $V_0$ one finds a corresponding 
$E_z$. Therefore, for each $V_0$, eqs. \ref{eq:sh11} determine a set 
$\sigma_2$, $\sigma_4$ and $R$, and these give  $N$ and other 
quantities such as the mean square radius 
 \begin{equation}                    
<r^2> = {{4 \pi R^4}\over {5N}} \left({\sigma_2\over 3} + 
{\sigma_4\over 7} \right)  \ ,
  \label{eq:sh12}
\end{equation}
or the various energies 
 \begin{eqnarray}                    
E_{con} &=& \int \  d^2r U_{2D,c}(r) \sigma_q(r) = \pi R^4 \bigg[ 
\sigma_2 \left( {2\over 15} {\bar k}_c + \right. \nonumber  \\ 
&+& \left. {8\over 105} {\bar k}_4 R^2 \right) 
 + \sigma_4 \left( {2\over 35} {\bar k}_c + {8\over 315} {\bar k}_4 
R^2\right) \bigg] \ , 
 \label{eq:sh13}
\end{eqnarray}
 \begin{eqnarray}                    
E_{Coul,d} &=& {1\over 2} \int \ d^2r U_{2D,d}(r)  \sigma_q(r) = 
{{\pi^3 e^2 R^3}\over{4 \varepsilon_r}} \nonumber \\ &.& 
\bigg[ \sigma_2^2 {4\over 5}\bigg({2\over 3}-{3\over 7}\tau_2  \bigg) 
+ \sigma_4^2 {1\over 35} \bigg(8 -{19\over 3} \tau_4 \bigg)  \nonumber \\ 
 &+&  \sigma_2 \sigma_4 {1\over 7}\bigg({24\over 5}-{16\over 9}\tau_2 
-{9\over 5} \tau_4  \bigg) \bigg]  \nonumber \\
E_{Coul,m} &=& {1\over 2} \int \ d^2r U_{2D,m}(r) \sigma_q(r) = 
-{{e^2 \pi N}\over{\varepsilon_r L}} R^2 \bigg[ \sigma_2 \bigg( {1\over 3} 
\nonumber \\ 
 &-&{R^2\over{15 L^2}} +{R^4\over{35 L^4}} \bigg) + \sigma_4 \left( 
{1\over 5} -{R^2\over{35 L^2}} + {R^4\over {105 L^4}} \right) \bigg] 
 \nonumber \\
E_{kin} &=& \int \  d^2r \ {{\hbar^2\pi}\over {2m^*}} 
\sigma_q^2(r) \nonumber \\ 
&=& {{\hbar^2 \pi^2}\over{2m^*}} R^2 \bigg[ {1\over 2} \sigma_2^2 
+ {2\over 3} \sigma_2 \sigma_4 + {1\over 4} \sigma_4^2\bigg]
\label{eq:sh14}
\end{eqnarray}

 \begin{figure}[htpb]                 
\leavevmode
\epsfxsize=8cm
\epsffile{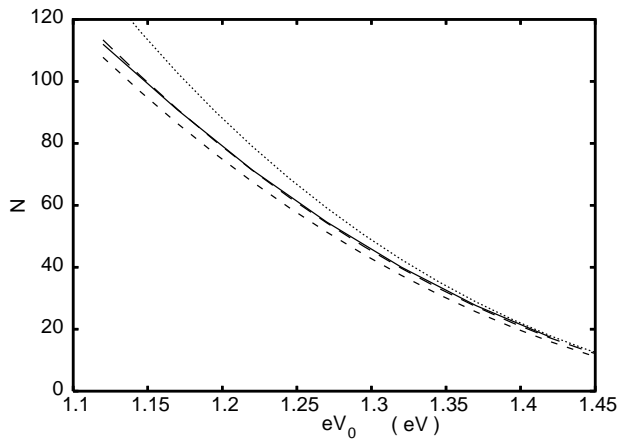}
\caption{Electron number $N$ obtained from the various models:
  continuous line: factorized Hartree. Long dashed line overlapping 
with it on most of the range: 
parametrized density, $\sigma_q(r)$, with parameters determined as 
described in Section V. Short dashed line underneath: same but 
omitting the non-zero thickness corrections in the coulomb terms 
($\tau_2 =\tau_4 =0$.) Upper dotted line: same but omitting the 
quartic term in the confining potential (${\bar k}_4=0$).   } 
 \label{figl06}
\end{figure} 

\begin{figure}[htpb]                  
\leavevmode
\epsfxsize=8cm
\epsffile{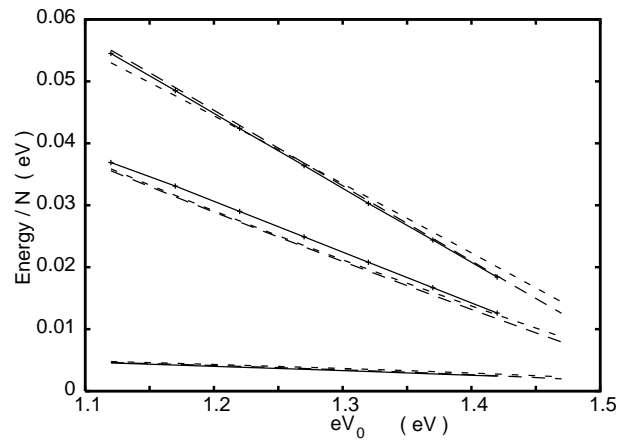}
\caption{Energies per electron v.s. gate voltage. From top to bottom: 
Coulomb, confining and kinetic. Continuous lines: 
Hartree with factorization ansatz. Long dashed lines: predictions of 
parametrized densities of eq. \protect{\ref{eq:sh3}} and expressions 
\protect{\ref{eq:sh13}} to \protect{\ref{eq:sh14}}. Short dashed lines: 
analytic approximations: eqs. \protect{\ref{eq:sh21}}. } 
 \label{figl07}
\end{figure}

\begin{figure}[htpb]                  
\leavevmode
\epsfxsize=8cm
\epsffile{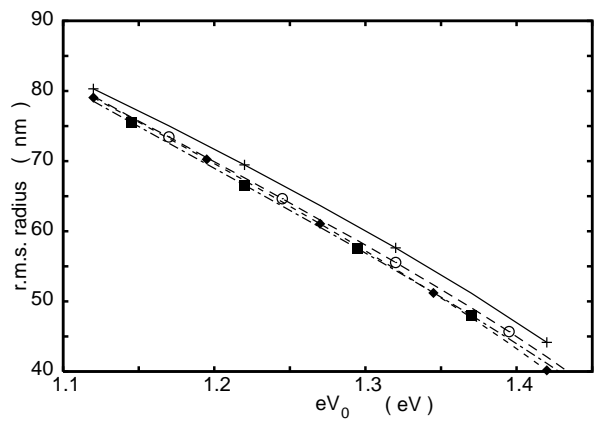}
\caption{ Root mean square radius v.s. gate voltage. Continuous line 
with ``+'' signs: Hartree, (factorization ansatz). Dashed line with circles: 
parametrized density with two terms.  Dotted line with diamonds: 
parametrized density with one term, eq. \protect{\ref{eq:sh18}}. 
Dash-dotted line with boxes: prediction of eq. 
\protect{\ref{eq:sh19}}. } 
 \label{figl08}
\end{figure}

\begin{figure}[htpb]                  
\leavevmode
\epsfxsize=8cm
\epsffile{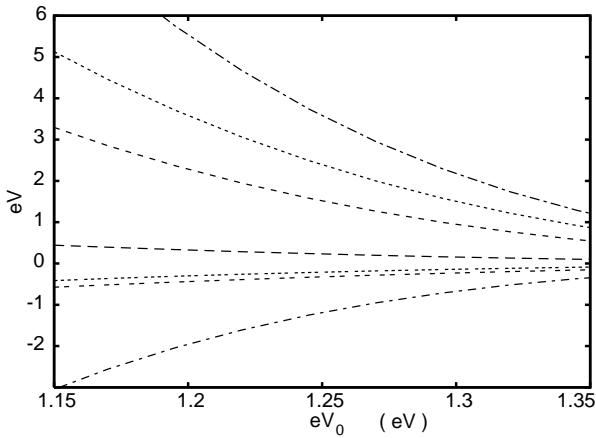}
\caption{ Energies relevant to the equilibrium configurations v.s. 
$V_0$: from top to bottom: Coulomb direct, Coulomb direct plus 
mirror, confining, kinetic, Coulomb finite thickness effect, 
Coulomb exchange and  Coulomb mirror terms.  } 
\label{figl10}
\end{figure}

\begin{figure}[htpb]                 
\leavevmode
\epsfxsize=8cm
\epsffile{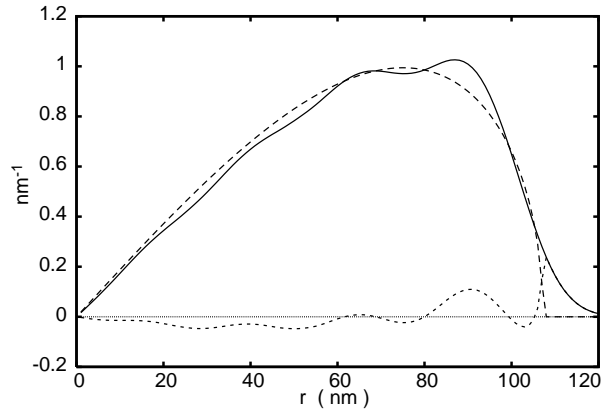}
\caption{ Radial distribution of charges: Densities weighted with a 
factor $2\pi r$, for $V_0 = -1.22 $ V. Continuous line: Hartree with 
factorization ansatz. Dashed line: parametrized density, eq. 
\protect{\ref{eq:sh3}}. Dotted line: difference.} 
\label{figl09}
\end{figure}

In Figs. \ref{figl06} to \ref{figl09} we compare the results 
of the present model to our Hartree calculations (with the 
separation ansatz). Fig. \ref{figl06} shows the number of electrons versus 
the gate potential. The prediction (long dashed line) 
is so close to the Hartree results that the two lines can only be 
distinguished when $eV_0$ is close to $1.1 $ eV. For comparison we 
also show the predictions when the finite thickness corrections, or the 
quartic terms, are omitted. As expected, the quartic terms are 
relevant only for large dots, whereas the finite size correction is 
sizeable even for small ones.  Fig. \ref{figl07}  shows the evolution 
of the various energies per particle: Coulomb (direct plus mirror), 
confining and kinetic, as $eV_0$ is varied. Again the agreement is 
excellent, confirming the validity of the model for bulk 
properties. The agreement is even better for the fields: for the 
example shown in Fig. \ref{figl04}, those given by the parametrized 
densities are so close that they cannot be distinguished from those 
plotted in the figure. This is a very useful result, since one can 
use them as starting values for self-consistent solution of the 
Hartree equations, and convergence is then very fast. 

Fig. \ref{figl08} shows the root mean square radii. Again the 
variation with $V_0$ is fairly well reproduced. But one observes a 
small systematic underestimate that we now examine for a specific 
case. Fig. \ref{figl09} compares the radial distributions of charge 
for a 
gate voltage of $V_0 = -1.22 $ V. As expected, the quantal 
oscillations of the Hartree density are more pronounced than
those from the parametrized 
density. However, the most 
significant disagreement occurs at the surface, where the form of 
$\sigma_q$ only allows an interpolation between the oscillations of 
$\sigma_H$. This is clearly shown by their difference, the dotted line 
in that figure. Still it is seen that most of the charge is 
correctly located, and this is why the energies 
and fields are so accurately  predicted by the model. 

Fig. \ref{figl10} is a more detailed plot of relevant energies v.s. 
$V_0$. We show  separately the contributions 
from the direct and mirror Coulomb terms. The mirror term is quite 
large and should not be ignored in any 2D calculation. We also show 
three comparatively small terms: the kinetic energy (positive), the 
Coulomb exchange (negative) and the correction to Coulomb due to the 
finite thickness of the 2DEG. The latter correction is obtained from the 
expression for the Coulomb energy in eq. \ref{eq:sh14} as  
 \begin{equation}                    
\Delta E_{Coul,d} = E_{Coul,d} - E_{Coul,d}( \tau_2 =0, \tau_4 =0) \ .
\label{eq:sh15}
\end{equation}

For comparison, the Coulomb exchange energy has also been computed 
following Zaremba \cite{Z96},  although the effect of the 
corresponding effective potential has not been included in our model. 
Notice that its contribution is comparable to the effect of the 
finite thickness on the Coulomb energy, a term that pure 2D models 
neglect. 

\subsection{Analytic approximations for the radii and bulk energies}  
The values of $\sigma_4$ are typically about ten percent of 
those of $\sigma_2$. Therefore, if one is willing to trade some 
accuracy for simplicity, $\sigma_4$ can be neglected. In this approximation, 
$\sigma_q(r)$, eq. \ref{eq:sh3}, is replaced by  
$\sigma_p(r)$ of eq. \ref{eq:sh2} and the direct coulomb potential reduces to: 
 \begin{equation}                   
U_{2D,d} = {{\pi^2 e^2}\over {2\varepsilon_r}} R \sigma_2 \left(1 - 
\tau_2 + {1\over 2}(3 \tau_2-1){r^2\over R^2} -{\tau_2\over 2} 
{r^4\over R^4} \right) \ ,
\label{eq:sh16}
\end{equation}
which, due to the smallness of $\tau_2$, is practically parabolic. It 
is clear then that imposing eq. \ref{eq:sh10} for all $r\le R$ will 
lead to unphysical results. What we do instead is to impose it {\it 
a)} as a weighted average over all $r$, with $\sigma_p(r)$ as 
weight function and  {\it b)} at $r=0$, where the quartic terms are 
not relevant. We then arrive at two equations: 
 \begin{eqnarray}                    
&-&E_z = {1\over 5} {\bar k}_c R^2 +{4\over 35} {\bar k}_4 R^4 + 
\sigma_2 \bigg[ {{3 \hbar^2\pi}\over {4 m^*}} \nonumber \\ 
 &+& {{3 \pi^2e^2}\over{5\varepsilon_r}} R \left({2\over 3}- {3\over 
7} \tau_2\right) - {{2\pi e^2}\over{\varepsilon_r}} {R^2\over L} 
\bigg({1\over 3} -{R^2\over {15 L^2}} +{R^4\over{35 L^4}} \bigg) 
\bigg] \nonumber \\ 
 &-&E_z = \sigma_2 \bigg( {{\hbar^2\pi}\over m^*} +{{\pi^2 e^2}\over 
{2\varepsilon_r}} R ( 1-\tau_2) -{{2\pi e^2}\over {3\varepsilon_r}} 
{R^2\over L} \bigg) 
 \label{eq:sh17}
\end{eqnarray}
which can again be reduced to a single non-linear equation for $R$: 
 \begin{eqnarray}                    
R^2 &=& 
-E_z 
{ {1- {17\over 7} \tau_2 - {4\over \pi} {R^3\over L^3} \left({1\over 3} - 
{R^2\over{7 L^2}} \right)+ 5 \xi_k }\over
{( {\bar k}_c + {4\over 7} {\bar k}_4 R^2)\, 
(1 -\tau_2 -{4\over {3\pi}} {R\over L} + 4 \xi_k) } } 
\label{eq:sh18}
\end{eqnarray}
where $\xi_k \equiv \hbar^2/(2m^*) \varepsilon_r/(\pi e^2 R) \simeq 
1.65/R $ is the small correction due to the kinetic energy. Similarly 
the finite thickness corrections are only a few percent. Note also 
that in both cases there is a significant cancellation between the 
corrections in the numerator and in the denominator, so to a first 
approximation they cancel. By neglecting the mirror and 
the quartic term contributions one arrives at Shikin's simple result 
 \begin{equation}                    
R_S^2 = - {E_z\over {\bar k}_c}~. 
\label{eq:sh19}
\end{equation}
Due to other cancellations that we will discuss below, $R_S$ is 
fairly close to the value of $R$ found by solving eq. \ref{eq:sh18} 
without simplification. This is shown in Fig. \ref{figl08}. To give a 
specific example, when $V_0 = -1.27 $ V, the value of $R$ from eq. 
\ref{eq:sh11} is $ 98.0 $ nm, the value from eq. \ref{eq:sh18} is 
$96.5 $ nm, and from eq. \ref{eq:sh19} $R_S = 95.8 $ nm. One can thus 
find a good approximate solution for $R$ by inserting $R_S$ for $R$ 
on the rhs of eq. \ref{eq:sh18}. This gives $R= 96.6 $ nm, and on the 
scale of Fig. \ref{figl08} the exact and the approximate solutions to 
eq. \ref{eq:sh18} are indistinguishable. The agreement with the 
Hartree results is also quite good. This clearly shows that eqs. 
\ref{eq:sh18} and \ref{eq:sh19} can be used to quantify the effect of 
the different terms on the size of the dot.

Expanding the second denominator in eq. \ref{eq:sh18} and neglecting 
small terms gives 
 \begin{equation}                    
R^2 \approx R_S^2 \left( 1 - {10\over 7} \tau_2 +{4\over{3\pi}} 
{R_S\over L_S} -{4\over 7} {{{\bar k}_4 R_S^2}\over{{\bar k}_c}} 
 + \xi_k\right) \ ,  
 \label{eq:sh20}
\end{equation}
which is less accurate, particularly for large dots. But, the usefulness 
of this simplified expression is that it displays very simply the expected 
role of each contribution: the finite 
thickness correction and the confining quartic term decrease the 
radius, whereas the mirror term increases it. For our chosen 
heterostructure, cancellations between the contributions to eq. 
\ref{eq:sh20} leave the final dot size very close to $R_S$. The 
dependence of $R_S$ on $V_0$ can be brought out using 
analytic expressions from the appendices: the numerator of eq. 
\ref{eq:sh19}, $E_z$, can be approximated by eq. \ref{eq:ap2}, 
whereas the denominator is linear in $V_0$ as shown in eq. 
\ref{eq:bp5}. From these expressions, the predicted values of $R_S$ 
fall between the continuous line in Fig. \ref{figl08} describing the 
Hartree results and the dashed line corresponding to solutions of eq. 
\ref{eq:sh18}. Summarizing: due to mutually 
compensating effects, the main trends in the dependence of the dot 
size on the gate voltage can be easily understood with the help of 
the expressions given in the appendices for the numerator and denominator 
of eq. \ref{eq:sh19}. 
A similar analysis can be made for the corrections to $\sigma_2$ 
using the second of eqs. \ref{eq:sh17}. Then from eqs. \ref{eq:sh13} and 
\ref{eq:sh14} we find 
 \begin{eqnarray}                      
{{E_{Coul}}\over N} &=& -E_z \Bigg[ {2\over 5}{{\displaystyle{1- 
{9\over14} \tau_2}}\over{\displaystyle{1-\tau_2 -
{4\over{3\pi}}{R\over L}+ 2 {\hbar^2\over m^*} 
{\varepsilon_r\over{\pi e^2 R}}}}} \nonumber \\ && \hskip 2.7cm
 -{2\over{3\pi}} {R\over L} {{\displaystyle{1-{R^2\over{5L^2}}+ 
{R^4\over {7L^4}}}}\over{\displaystyle1-\tau_2 -{{4R}\over {3\pi 
L}}}} \Bigg] \nonumber \\ 
 {{E_{con}}\over N} &=& {R^2\over 5} \left({\bar k}_c +{4\over 7} 
{\bar k}_4 R^2 \right) \nonumber \\ 
 {{E_{kin}}\over N} &=& -E_z{{\displaystyle{{3\over 8}{{\hbar^2\pi }\over 
m^*}}}\over{\displaystyle{{{\hbar^2\pi}\over m^*} 
+{{\pi^2e^2}\over {2\varepsilon_r}} R ( 1-\tau_2 ) - {{2\pi e^2}\over 
{3\varepsilon_r}}{R^2\over L} }}} \ . 
 \label{eq:sh21}
\end{eqnarray}
Setting $R = R_S$ and using the analytic expression for $E_z$, 
eq. \ref{eq:ap2}, the predictions are as shown in Fig. \ref{figl07}: 
they are quite accurate, so that one can see by inspection of 
eqs. \ref{eq:sh21} the role of each term in determining the bulk 
energies given by the 3D Hartree calculations.

\section{ Summary and conclusions} 

Starting from a 3D Poisson-Schr\"odinger  model, we have studied a 
sequence of approximations. The 3D Hartree results show that although 
the electrons are confined around the junction plane, they are spread 
longitudinally more than $20 $ nm into the substrate. We have 
introduced a factorization {\it ansatz} that is very accurate, by 
factorizing out a common longitudinal component of the wavefunctions, 
as a first step towards pure 2D models. 
  In the second half of the manuscript we have focussed on improving 
the semi-classical model of Shikin {\it et al}. \cite{Shik91}. We 
have formulated a variational approach where in addition to the 
classical ingredients, (confinement and direct Coulomb potentials), 
we take into account the kinetic energy via the Thomas Fermi 
approximation, the mirror coulomb terms and the finite thickness of 
the 2DEG. We have shown that using parametrized densities in the 
variational equations leads to good values for the number of electrons, 
mean square radius, kinetic, confining and Coulomb energies. 
Furthermore, the corresponding total potential seen by the 
electrons is so close to that found in the Hartree factorization 
{\it ansatz}, that it provides a very efficient starting point for 
full Hartree calculations. In an additional  simplifying step  we 
have shown that choosing the simplest parametrization for the 
density, but keeping all potential contributions, the dependence of 
the bulk properties on the gate voltage can be described 
analytically. These expressions clarify the role of each parameter in 
determining the dot properties.  This detailed and 
quantitative connection between 3D simulations and the simplest 2D 
models provides an essential tool for discussing the systematics 
of bulk quantum dot properties in a transparent and intuitive way. 

\acknowledgments
We are grateful to NSERC-Canada for Discovery Grant RGPIN-3198 
(DWLS), and to DGES-Spain for continued support through grants  
FIS2004-03156 and FIS2006-10268-C03-02 (JM). 


\appendix

\section{ Approximate longitudinal wavefunctions}

\subsection{Airy functions} 
\begin{figure}[htpb]                   
\leavevmode 
\epsfxsize=8cm 
\epsffile{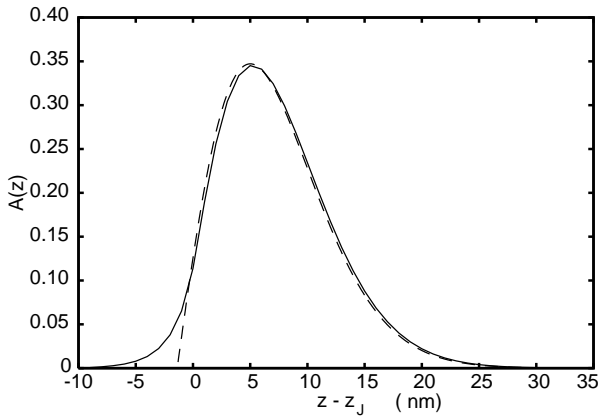} 
\caption{Longitudinal wavefunction as determined using the 
factorization ansatz (continuous line), compared to the Airy 
approximation described in Appendix A (dashed line.) $V_0 = -1.27$ 
V.} 
\label{figl11} 
\end{figure}

Here we describe how we approximate the longitudinal wavefunctions 
$A(z)$, by Airy functions.  It is well known that the potential term 
$eV_{g+d}(0,z)$  is close to linear in the substrate, 
and a steep barrier in the spacer layer.  Therefore, in the substrate 
we approximate eq. \ref{eq:le19a} by
 \begin{equation}                       
- {\hbar^2 \over {2m^*}} A_l''(z) + \gamma (z-z_J) A_l(z) = E_l A_l(z)
\label{eq:ap1}
\end{equation}
where $z_J$ is the location of the junction between substrate and 
spacer, and $\gamma$ a constant slope whose value will be fixed 
below. In this approximation: 
 \begin{eqnarray}                       
E_z &=& E_l + eV_s -{{4\pi e^2}\over \varepsilon_r} \rho_d d ( c+{d\over 2} )
+ eV_0 { z_J\over{\sqrt{z_J^2+S_0^2}}} \ .
\nonumber \\ 
\label{eq:ap2}
\end{eqnarray}
Introducing the change of variables 
 \begin{equation}                       
u = \alpha (z-z_J) + \xi_0~, 
\label{eq:ap3}
\end{equation}
with the identifications 
 \begin{eqnarray}                       
\alpha &=& \left( {{2m^* \gamma} \over \hbar^2}\right)^{1/3} \nonumber \\
E_l &=& -\xi_0 \gamma^{2/3} \left( {\hbar^2 \over {2m^*}}
\right)^{1/3} \ ,
\label{eq:ap4}
\end{eqnarray}                          
eq. \ref{eq:ap1} reduces to the Airy equation in the form given on 
page 446 of Abramowitz and Stegun \cite{AS64} 
 \begin{equation}                       
A_l''(u) - u A_l(u) = 0 \ .
\label{eq:ap5}
\end{equation}
In the limit of an infinitely steep spacer barrier, $A_l(z)$ must 
vanish at $z=z_J$ so the first zero of the Airy function $ Ai(u)$ 
must occur at this point. Since the location of that zero is at 
$\xi_0 = - 2.33810$, the form of $A_l(z)$ is completely determined. 
 \begin{equation}                       
A_l(z) = {\cal N} Ai\left(\alpha (z-z_J) + \xi_0 \right) \Theta(z-z_J)~.
\label{eq:ap6}
\end{equation}
The normalization, and the location of the maximum are 
also analytic 
 \begin{eqnarray}                       
{\cal N} &=& {\sqrt{\alpha} \over {Ai'(\xi_0)}} \nonumber \\
z_m &=& z_J + {1\over \alpha}( \xi'_0- \xi_0)  \ ,
\label{eq:ap7}
\end{eqnarray}
with $\xi'_0 = -1.01879$. To have a good approximation for the 
potential where the wavefunction is largest, we choose to define the 
slope $\gamma$ as 
 \begin{equation}                       
\gamma \equiv \left[ {{d(eV_{g+d}(0,z)) }\over{dz}} \right]_{z=z_m} = 
eV_0 {{S_0^2}\over{(z_m^2+ S_0^2)^{3/2}}} \ . 
 \label{eq:ap8}
\end{equation}
For a given $V_0$, and the  $A(z)$ determined by solving eq. 
\ref{eq:le19a}, we construct its analytic approximation by 
shifting $A_l(z)$ to make the maxima of the two coincide. In Fig. 
\ref{figl11} we compare the  shifted $A_l(z)$ to  the numerically 
determined $A(z)$ for a representative gate voltage. One sees that 
the $A_l$ forbid tunneling into the spacer, but that their shape in 
the substrate is correct. 

\subsection{Averages over $z$}
As will be shown in Appendix B,  eq. 
\ref{eq:bp5}, the confining potential can be 
expanded as a power series in $r^2$. To extract from it an effective 
2D confining potential 
using eq. \ref{eq:le16} one has to perform an average over $z$ with 
weight function $A^2(z)$. We approximate this 
average using the $A_l(z)$. For the quadratic term one has to 
determine: 
 \begin{equation}
{\bar k}_c \equiv eV_0 \ {{3 S_0^2}\over 2} \int {z  \over{(z^2+ 
S_0^2)^{5/2}}} \ A^2(z) \ dz  \ , 
 \label{eq:ap9}
\end{equation}
and in this expression the denominator varies little ($z^2 << S_0^2$) 
in the range where $A(z)$ is nonvanishing. We remove it from the 
integral, leaving an average 
 \begin{equation}
\int z A^2(z) \ dz \simeq z_J + \int (z-z_J) A_l^2(z) \ dz = z_J - 
{2\over 3} {\xi_0 \over \alpha} \ . 
 \label{eq:ap10}
\end{equation}
Combining this approximation with the second of 
eqs. \ref{eq:ap7} we find: 
 \begin{equation}
{\bar k}_c = k_c({\bar z})= {3\over 2} eV_0 \ {{{\bar z} S_0^2}\over {({\bar z}^2+S_0^2)^{5/2}}} \ ,
\label{eq:ap11}
\end{equation}
with
 \begin{equation}                       
{\bar z} = z_m -{{\xi'_0-\xi_0}\over \alpha} - {2\over 3} {\xi_0\over 
\alpha} \ , 
 \label{eq:ap12}
\end{equation}
where $z_m$ is the location of the maximum of $A(z)$. The quartic 
term in the confining potential can also be approximated with the 
same prescription, neglecting the variation with $z$ in the 
denominator of the $r^4$ term in eq. \ref{eq:bp5} and keeping only the  
dependence on $z$ in the numerator. 
\begin{figure}[htpb]                    
\leavevmode
\epsfxsize=8cm
\epsffile{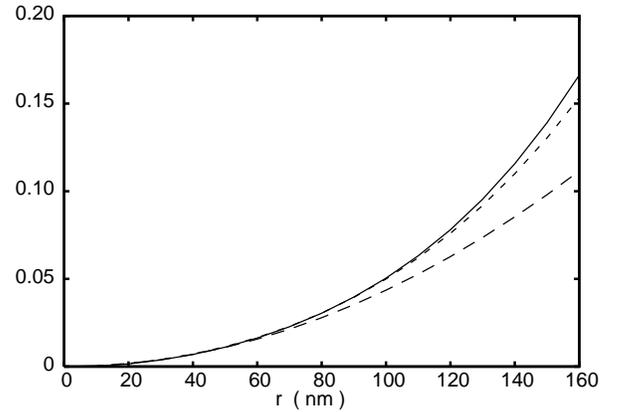}
\caption{ Confining potential $(V_g(r,z)-V_g(0,z))/V_0$ (solid line) 
compared to the power series expansion in $r^2$ with one (long dashed 
line)  and two terms (short dashed line ). $ z= 70 $ nm in this 
example.} 
\label{figl12} 
\end{figure}

\section{ Power series expansion of the gate potential} 
From eqs. \ref{eq:le1} and \ref{eq:le2}, when $z>0$ we have 
 \begin{equation}                       
eV_g ( r,z) \equiv {{eV_0}\over {2\pi}} \int_{S_0}^{\infty} s \ ds \ 
\int_0^{2\pi} \ d\theta \ { z \over{(z^2 +|{\vec r}- {\vec 
s}|^2)^{3/2}}}  \ , 
 \label{eq:bp1}
\end{equation}
To determine the series expansion in powers of $r$ for fixed $z$, we
use an auxiliary expansion that generates the derivatives $\partial^n
(eV_g )/ \partial r^n$ in a straightforward way.  We expand  
 \begin{eqnarray}                        
\bigg[z^2 &+&|{\vec r}-{\vec s}|^2\bigg]^{-3/2} = \left[ z^2 + s^2 -
2rs \cos \theta + r^2\right]^{-3/2} \nonumber \\ &=& (z^2+ s^2)^{-
3/2} \left[ 1- {{2rs \cos \theta -r^2}\over{z^2+s^2}}\right]^{-3/2} 
\nonumber \\ 
 &\simeq& (z^2+s^2)^{-3/2} \bigg( 1 + {3\over 2} \ {{2rs \cos \theta - 
r^2}\over {z^2+s^2}} \nonumber \\ &+& {15\over 8} \bigg({{2rs\cos 
\theta -r^2}\over{z^2+s^2}} \bigg)^2 + ... \bigg) \ , 
 \label{eq:bp2}
\end{eqnarray}
and integrate over $\theta$ term by term. To illustrate the steps we 
write only the first three terms in the expansion. 
The integration over $\theta$ gives  
 \begin{eqnarray}
eV_g(r, z) &=& e V_0 z                   
\int_{S_0}^{\infty} \ s \ ds \bigg[ {1\over{(z^2+s^2)^{3/2} }} 
-{3\over 2} {r^2\over{(z^2+s^2)^{5/2}}} \nonumber \\ &+& {15\over 4} 
\ {{r^2 s^2}\over{(z^2+s^2)^{7/2} }} + {15\over 8} \ 
{r^4\over{(z^2+s^2)^{7/2}}}\bigg] \ . 
 \label{eq:bp3}
\end{eqnarray} 
These integrals are analytical, so that 
 \begin{eqnarray}                        
eV_g(r,z) &=&  eV_0 \bigg[{z\over{\sqrt{z^2+S_0^2}}} \ + \ {3\over 4} {{z 
S_0^2}\over{(z^2+S_0^2)^{5/2} }} \ r^2 \nonumber \\ &+& {3\over 8} 
{z\over{(z^2+S_0^2)^{5/2}}} \ r^4 \bigg] \nonumber \\ 
 \label{eq:bp4}
\end{eqnarray}
To get all the contributions of order $r^4$, one has to include two 
more terms in the expansion of eq. \ref{eq:bp2}. We skip the details 
and quote the full result 
 \begin{eqnarray}                        
eV_g(r,z) &=& eV_0 \bigg[{z\over {(z^2 +S_0^2)^{1/2} }} + {3\over 4} 
{{zS_0^2}\over {(z^2+  S_0^2)^{5/2} }} \ r^2 \nonumber \\
&+& {15\over 64} {{z S_0^2 (-4 z^2+ 3S_0^2)}\over{(z^2+ S_0^2)^{9/2}}} 
\ r^4 + ... \bigg]  \ ,
\label{eq:bp5}
\end{eqnarray}
In Fig. \ref{figl12} we show the various results for $(V_g(r,z)- 
V_g(0,z))/V_0$ for $z= z_J = 70 $ nm, as a function of $r$.  As can 
be seen, the power series converges nicely towards the exact result. 
We have checked that similar convergence holds for other 
values of $z$. 

\section{ Finite thickness corrections} 

Here we derive the effective Coulomb potential to be used when 
treating the dot as a 2DEG. 
We first compute the exact correction at $r=0$ for the 
two terms of the parametrized density, eq. \ref{eq:sh3}. Then 
we show that to a good approximation a simple parametrization 
based on this correction can be used when $r \ne 0$. 

\subsection{ Potentials at $r=0$}                        
Using the effective interaction defined in eq. \ref{eq:le27}, and the 
first term of the parametrized density, the potential that we wish to 
compute requires the integral 
 \begin{equation}                        
I(t) = 2 \int_0^R \ {{\sqrt{1-r^2/R^2}}\over\sqrt{r^2 + t^2}} \ r \ dr \ .
\label{eq:cp1}
\end{equation}
With the change of variable: $ x = r^2$, the integral takes  a form 
that can be easily evaluated: 
 \begin{equation}                        
I(t) = -t + R \left( 1 + {t^2\over R^2}\right) \left( {\pi \over 2} -
\arctan {t\over R}\right) \ . 
 \label{eq:cp2}
\end{equation}
and when $t^2/R^2 << 1$  
 \begin{equation}                        
I(t)  \simeq -t + R \left({\pi\over 2} -{t\over R} \right) = {{\pi 
R}\over 2} - 2 t   \ ,
 \label{eq:cp3}
\end{equation}
so that defining 
 \begin{equation}                        
 \tau_2 = {{4t}\over {\pi R}}
\label{eq:cp4}
\end{equation}
one can finally write the finite thickness correction as $I(t)= 
(1- \tau_2) I(t=0)$. We have checked numerically  that indeed this 
approximation is quite good for the usual values of $R$ and the value 
of $t$ given in  eq. \ref{eq:le27}.\\ \\ 
For the second term of $\sigma_q$, eq. \ref{eq:sh3}, the integral of 
interest is 
 \begin{eqnarray}                        
I_q(t) &=& 2 \int_0^R {{(1- r^2/R^2)^{3/2}}\over{\sqrt{r^2+ t^2}}} \ r\ dr = 
\nonumber \\
&=& \int_0^{R^2} {{(1-x/R^2)^{3/2}}\over{\sqrt{x+t^2} }}\ dx \nonumber \\
&=& \int_0^{R^2} \ {{(1-x/R^2)^2  \ dx}\over{\sqrt{(x+t^2)(1-x/R^2)}}} \ ,
\label{eq:cp5}
\end{eqnarray}
Making use of integrals found on page 83 of \cite{Grad}, 
we arrive at  
 \begin{eqnarray}                        
I_q(t) &=& {3\over 8} R \left({\pi\over 2} + \arcsin {{R^2-
t^2}\over{R^2+t^2}}\right)\left(1+ {t^2\over R^2}\right)^2 \nonumber \\ 
&-& t \left({5\over 4} +{{3t^2}\over{4R^2}} \right) \ .
\label{eq:cp6}
\end{eqnarray}
This result reduces to $I_q(t=0) =3 \pi R/8$ in the limit $t=0$, for which the
integration is trivial. Then we define $\zeta_4 \equiv I_q(t)/I_q(0)$, and 
write
 \begin{eqnarray}                        
\zeta_4 &=& {8\over{3\pi}} \bigg[ {3\over 8}\left( {\pi\over 2} + 
\arcsin {{R^2-t^2}\over{R^2+t^2}}\right) \nonumber \\
&.& \left(1 + {t^2\over R^2}\right)^2 -{t\over{4R}} \left( 5 + 3 {t^2\over
    R^2}\right) \bigg] \ .
\label{eq:cp7}
\end{eqnarray}
If $t^2/R^2 << 1$ can be neglected, the expression for 
$\zeta_4$ simplifies to 
 \begin{equation}                        
\zeta_4 \equiv 1 - \tau_4 =  1-{10\over{3\pi}} \ {t\over R} \ .
\label{eq:cp8}
\end{equation}

\begin{figure} [htpb]                  
\leavevmode
\epsfxsize=8cm
\epsffile{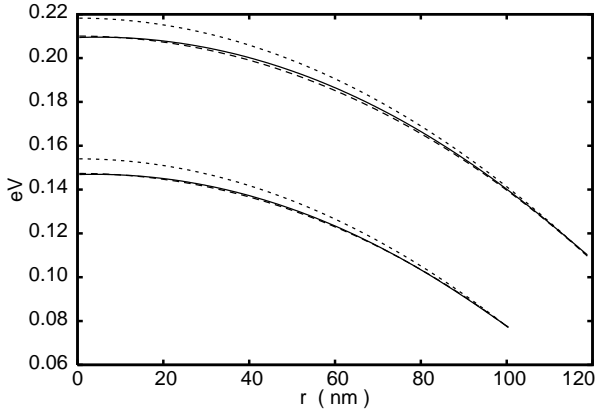}
\caption{ Direct Coulomb potential for the first term of the 
parametrized densities $\sigma_q(r)$, eq. \protect{\ref{eq:sh3}}. Values of 
$\sigma_2$ corresponding to $V_0 = -1.17 $ V (top) and $V_0 = -1.27 
$ V (bottom.) Continuous lines: exact numerical calculation using 
eqs. \protect{\ref{eq:le25}} and \protect{\ref{eq:tf5}}. 
Dotted lines: zero thickness eq. \protect{\ref{eq:cp9}}; dashed lines: 
including the finite thickness correction: eq. \protect{\ref{eq:cp10}}. } 
 \label{figl13}
\end{figure} 

\subsection{The direct Coulomb potential when $r \ne 0$}          
When the 2D interaction is taken as  $1/r$ the Coulomb 
potential for the density $\sigma_p(r)$ is analytic: 
  \begin{equation}
U_{2D, d} (r) = {{\pi^2 e^2}\over{2\varepsilon_r}} R  \sigma_2 
\left( 1- {r^2\over{ 2R^2}}\right) \ . 
 \label{eq:cp9}
\end{equation}
In Fig. \ref{figl13} we compare this curve, dotted lines, to the 
exact numerical calculation using eqs. \ref{eq:le25} and 
\ref{eq:tf5}. As expected there is a sizeable discrepancy where the 
dot charge is large, and as the distance from the center of the dot 
increases the thickness effect decreases. The dashed lines are our 
empirical prescription for extending the finite thickness correction 
to non vanishing $r$: 
 \begin{eqnarray}
U_{2D, d}^{(2)} (r) &=& {{\pi^2 e^2}\over{2\varepsilon_r}} R  \sigma_2
\left( 1- {r^2\over{ 2R^2}}\right) \left( 1- \tau_2 \left( 1- {r^2\over
      R^2}\right) \right) \nonumber \\ 
\label{eq:cp10}
\end{eqnarray}
As shown in Fig. \ref{figl13}, this prescription accurately 
reproduces the exact calculations.  
For the small contribution of the second term in the parametrized 
density we use a similar prescription: 
 \begin{eqnarray}
U_{2D, d}^{(4)} (r) &=& {{3\pi^2 e^2}\over{8 \varepsilon_r}} R \sigma_4
\left( 1-{r^2\over R^2}+{3\over 8} {r^4\over R^4} \right) \nonumber 
\\ && \quad \times \left( 1- \tau_4 \left( 1- {r^2\over R^2}\right) \right)~, 
 \label{eq:cp11}
\end{eqnarray}
and the results (not shown) are also sufficiently accurate.


\noindent file: lepdot16.tex  February 15, 2007


\begin{thebibliography}{44}

\bibitem[1]{CSS00} R. Crook, C.G. Smith, M.Y. Simmons and D.A. 
Ritchie, J. Phys.: Cond. Matter {\bf 12} (2000) L735-L740. 

\bibitem[2]{CST02} R. Crook, C.G. Smith, W.R. Tribe, S.J. O'Shea, 
M.Y. Simmons and D.A. Ritchie, Phys. Rev. B {\bf 66} (2002) 121301(R). 

\bibitem[3]{CSG03} R. Crook, C.G. Smith, A.C. Graham, I. Farrer, 
H.E. Beere and D.A. Ritchie, Phys. Rev. Lett. {\bf 91} (2003) 246803. 

\bibitem[4]{PKI04} A. Pioda, S. Ki$\check{c}$in, T. Ihn, M. Sigrist, A. Fuhrer,
K. Ensslin, A. Weichselbaum, S.E. Ulloa, M. Reinwald and W. Wegscheider,
Phys. Rev. Lett. {\bf 93} (2004) 216801. 

\bibitem[5]{KPI05} S. Ki$\check{c}$in, A. Pioda, T. Ihn, M. Sigrist, A. Fuhrer, 
K. Ensslin, M. Reinwald and W. Wegscheider, New J. Phys. {\bf 7} (2005) 185. 

\bibitem[6]{Kumar90} A. Kumar, S. E. Laux and F. Stern, 
Phys. Rev. B {\bf 42} (1990) 5166. 

\bibitem[7]{JL94} D. Jovanovic and J.P. Leburton, Phys. Rev. B {\bf 
49} (1994) 7474. 

\bibitem[8]{Sto96} M. Stopa, Phys. Rev. B {\bf 54} (1996) 13767. 

\bibitem[9]{GMI99} M. Governale, M. Macucci, G. Iannaccone, C. 
Ungarelli and J. Martorell, J. Appl. Phys. {\bf 85} (1999) 2962. 

\bibitem[10]{MHI93} M. Macucci, K. Hess and G.J. Iafrate, 
Phys. Rev. B {\bf 48} (1993) 17354. 

\bibitem[11]{FV94} M. Ferconi and G. Vignale, 
Phys. Rev. B {\bf 50} (1994) 14722. 

\bibitem[12]{WW95} Y. Wang, J. Wang, H. Guo and E. Zaremba, Phys. 
Rev. B {\bf 52} (1995) 2738. 

\bibitem[13]{JKW97} L. Jacak, J. Krasnyj and A. Wojs, Physica B {\bf 
229} (1997) 279. 

\bibitem[14]{HW99} K. Hirose and N.S. Wingreen, Phys. Rev. B {\bf 
59} (1999) 4604. 

\bibitem[15]{RM02} S.M. Reimann and M. Manninen, Rev. Mod. Phys. 
{\bf 74} (2002) 1283. 

\bibitem[16]{Shik89} V. Shikin, T. Demel and D. Heitmann, 
Sov. Phys. JETP {\bf 69} (1989) 797-803.

\bibitem[17]{Shik91} V. Shikin,, S. Nazin, D. Heitmann and T.
Demel, Phys. Rev. B {\bf 43} (1991) 11903-7.

\bibitem[18]{YZ94} Z.L. Ye and E. Zaremba, Phys. Rev. B {\bf 50} (1994) 17217.

\bibitem[19]{MWS94} J. Martorell, Hua Wu and D.W.L. Sprung, 
Phys. Rev. B {\bf 50 } (1994) 17298-308. 

\bibitem[20]{Laux88} S. E. Laux, J. D. Franck and F. Stern, 
Surf. Sci.  {\bf 196} (1988) 101.

\bibitem[21]{Davies88} J. H. Davies, 
Semicond. Sci. Technol. {\bf 3} (1988) 995. 

\bibitem[22]{LDK93} A.R. Long, J.H. Davies, M. Kinsler, S. Vallis and 
M.C. Holland, Semicond. Sci. Technol. {\bf 8} (1993) 1581. 

\bibitem[23]{JUYB04} Hong Jiang, D. Ullmo, Weitao Yang and H.U. Baranger, 
Phys. Rev. B {\bf 69} (2004) 235326.

\bibitem[24]{YBB01} I.I. Yakimenko, A.M. Bychkov and K.F. Berggren, 
Phys. Rev. B {\bf 63} (2001) 165309.

\bibitem[25]{DLS95} J.H. Davies, I.A. Larkin and E.V. Sukhorukov, J. 
Appl. Phys. {\bf 77} (1995) 4504. 

\bibitem[26]{RD03} O.E. Raichev and P. Debray, 
J. Appl. Phys. {\bf 93} (2003) 5422.

\bibitem[27]{PIK02} M.G. Pala, G. Iannaccone, S. Kaiser, A. Schliemann,
  L. Worschech and A. Forchel, Nanotechnology {\bf 13} (2002) 373.

\bibitem[28]{FIM02} G. Fiori, G. Iannaccone, M. Macucci, S. Reizenstein, 
S. Kaiser, M. Kesselring, L. Worschech and A. Forchel, Nanotechnology {\bf
  13} (2002) 299.

\bibitem[29]{AS64} M. Abramowitz and I.A. Stegun, \lq\lq Handbook
of Mathematical Functions", Dover Publications (New York), 
(1965). 

\bibitem[30]{SM96} D.W.L. Sprung and J. Martorell, 
Sol. State Comm. {\bf 99} (1996) 701-6. 

\bibitem[31]{Z96} E. Zaremba, Phys. Rev. B {\bf 53} (1996) R10512.

\bibitem[32]{Grad} I.S. Gradshteyn and I.M. Ryzhik, ``Table of 
Integrals, Series and Products'', Academic Press 1980. 


\end{thebibliography}
\end{document}